\documentclass[12pt,a4paper]{article}
\usepackage{graphicx}
\usepackage{amsmath,amssymb}
\usepackage{amsfonts}
\begin{document}
\begin{center}

 {\bf STRUCTURE OF LOW-LYING QUADRUPOLE STATES \\IN NUCLEI NEAR $^{132}$Sn}

 \medskip
 A. P. Severyukhin$^{1}$, V. V. Voronov$^{1}$, N. N. Arsenyev$^{1}$,
 N. Pietralla$^{2}$,\\ Nguyen Van Giai$^{3}$

 \medskip
{\it
 $^{1}$Bogoliubov Laboratory of Theoretical Physics,
             JINR, 141980 Dubna, Russia \\
 $^{2}$Institut f\"{u}r Kernphysik, Technische Universit\"{a}t
  Darmstadt, 64289 Darmstadt, Germany\\
 $^{3}$Institut de Physique Nucl\'eaire, CNRS-IN2P3,
Universit\'e Paris-Sud, F-91406 Orsay Cedex, France }
\end{center}

\vspace*{-1cm}
\begin{abstract}
The properties of the low-lying $2^+$ states in the even-even
nuclei around $^{132}$Sn are studied within the quasiparticle
random phase approximation. Starting from a Skyrme interaction in
the particle-hole channel and a density-dependent zero-range
interaction in the particle-particle channel, we use the finite
rank separable approach in our investigation. It is found that the
$2^+_4$ state in $^{132}$Te could be a good candidate for a
mixed-symmetry state.
\end{abstract}

\bigskip
\section{Introduction}

The low-energy spectrum of nuclear excitations provides the most
sensitive testing ground for nuclear structure calculations in the
presence of pairing correlations. The evolution of the low energy
spectrum in nuclei around $^{132}$Sn is an increasingly important
point for the investigation of nuclear structure physics and
nuclear astrophysics. There is a relation between the $N=82$ shell
closure and the $A\approx130$ peak of the solar r-process
abundance distribution, i.e., the structural peculiarities of the
$N=82$ isotones lighter than $^{132}$Sn are important for the
stellar nucleosynthesis. New
experiments~\cite{Rad02,Jak02,Scher04,Rad05,Gar06,Jun07,Dwo08,pbl08,cd128}
give spectroscopic observations in nuclei near $^{132}$Sn and this
is a good possibility to test theoretical approaches. One of the
successful tools for nuclear structure studies is the
quasiparticle random phase approximation (QRPA) with the
self-consistent mean-field derived by making use of the Skyrme
effective nucleon-nucleon interaction~\cite{vau72}. Such QRPA
calculations do not require to introduce new parameters since the
residual interaction is derived from the same energy density
functional as that determining the mean-field. Results of
calculations of the properties of the low-lying states in nuclei
with very different mass numbers within this approach are in a
reasonable agreement with experimental data.

Making use of the finite rank separable
approximation~\cite{gsv98,ssvg02} for the residual interaction
enables one to perform the QRPA calculations in very large
two-quasiparticle spaces. Recently, we generalized our approach to
take into account a coupling between the one- and two-phonon
components of the wave functions~\cite{svg04}. Here, we use an
extension of our approach by taking into account the
particle-particle (p-p) residual interaction~\cite{svg08}.

In the present report we briefly describe our method and present
the analysis of the properties of the low-lying $2^+$ states in
even-even nuclei around $^{132}$Sn.
%
%
\section{The method}
This method has already been presented in details in
Refs.~\cite{gsv98,ssvg02,svg08} and we only sketch it briefly
here. The starting point of the method  is the HF-BCS
calculation~\cite{RingSchuck} of the ground states, where
spherical symmetry is imposed on the quasiparticle wave functions.
The continuous part of the single-particle spectrum is discretized
by diagonalizing the HF Hamiltonian on a harmonic oscillator
basis~\cite{bg77}. We work in the quasiparticle  representation
defined by the canonical Bogoliubov transformation:
\begin{equation}
a_{jm}^{+}\,=\,u_j\alpha _{jm}^{+}\,+\,(-1)^{j-m}v_j\alpha _{j-m},
\end{equation}
where $jm$ denote the quantum numbers $nljm$. The Skyrme
interaction~\cite{vau72} is used in the p-h channel while the
interaction in the particle-particle (p-p) channel is assumed to
be a surface peaked density-dependent zero-range force:
\begin{equation}
V_{pair}({\bf r}_1,{\bf r}_2)=V_{0}\left( 1-\frac{\rho \left(
r_{1}\right) }{\rho _{c}}\right) \delta \left( {\bf r}_{1}-{\bf
r}_{2}\right) \label{pair}
\end{equation}
The strength $V_{0}$ is a parameter which is fixed to reproduce
the odd-even mass difference of nuclei in the studied
region~\cite{svg08}.

The residual interaction in the p-h channel $V^{ph}_{res}$ and in
the p-p channel $V^{pp}_{res}$ can be obtained as the second
derivative of the energy density functional with respect to the
particle density $\rho$ and the pair density $\tilde{\rho}$,
accordingly. Hereafter we simplify $V^{ph}_{res}$ by approximating
it by its Landau-Migdal form. For Skyrme interactions all Landau
parameters with $l > 1$ are zero. We keep only the $l=0$ terms in
$V^{ph}_{res}$. In this work we study only normal parity states
and one can neglect the spin-spin terms since they play a minor
role~\cite{ssvg02}. The two-body Coulomb and spin-orbit residual
interactions are also dropped. Therefore we can write the residual
interaction in the following form:
\begin{eqnarray}
V^{a}_{res}({\bf r}_1,{\bf r}_2) &=&N_0^{-1}[ F_0^{a}(r_1)+
F_0^{'a}(r_1)({\bf \tau }_1\cdot{\bf \tau }_2)] \delta ({\bf r}_1
- {\bf r}_2),
\end{eqnarray}
where $a$ is the channel index $a=\{ph,pp\}$; $N_0 =
2k_Fm^{*}/\pi^2\hbar^2$. The expressions for $F^{ph}_0, F^{'ph}_0$
and $F^{pp}_0, F^{'pp}_0$ can be found in Ref.\cite{sg81} and in
Ref.\cite{svg08}, respectively.

The p-h matrix elements and the antisymmetrized p-p matrix
elements can be written as a sum of separable terms in the radial
coordinate~\cite{gsv98,ssvg02,svg08}. Indeed, after integrating
over the angular variables one has to calculate the radial
integrals,
\begin{eqnarray}
\label{I} I^{a}(j_{1}j_{2}j_{3}j_{4})
&=&N_{0}^{-1}\int_{0}^{\infty }\left( F_{0}^{a}(r)+F_{0}^{'
a}(r)\mathbf{\tau }_{1}\cdot \mathbf{\tau }
_{2}\right)  \nonumber\\
&&\times
u_{j_{1}}(r)u_{j_{2}}(r)u_{j_{3}}(r)u_{j_{4}}(r)\frac{dr}{r^{2}}
\end{eqnarray}
where $u_{j}(r)$ is the radial part of the single-particle wave
function. The radial integrals~(\ref{I}) can be calculated
accurately by choosing a large enough cut-off radius and using a
$N$-point integration Gauss formula. Thus, the residual
interaction can be expressed as a sum of $N$ separable terms. The
Hamiltonian of our method has the same form as the Hamiltonian of
the well-known quasiparticle-phonon model~\cite{solo}, but the
single-quasiparticle spectrum and the parameters of the residual
interaction are calculated from the Skyrme forces.

We introduce the phonon creation operators
\begin{equation}
Q_{\lambda \mu i}^{+}\,=\,\frac 12\sum_{jj^{^{\prime }}}\left( X
_{jj^{^{\prime }}}^{\lambda i}\,A^{+}(jj^{^{\prime }};\lambda \mu
)-(-1)^{\lambda -\mu }Y _{jj^{^{\prime }}}^{\lambda
i}\,A(jj^{^{\prime }};\lambda -\mu )\right),
\end{equation}
The index $\lambda $ denotes total angular momentum and $\mu $ is
its z-projection in the laboratory system. One assumes that the
ground state is the phonon vacuum $| 0 \rangle $. We define the
excited states as $Q_{\lambda\mu i}^{+} | 0\rangle$. Making use
of the linearized equation-of-motion approach one can get the QRPA
equations~\cite{RingSchuck}:
\begin{equation}
\label{QRPA} \left(
\begin{tabular}{ll}
$\mathcal{A}$ & $\mathcal{B}$ \\
$- \mathcal{B}$ & $- \mathcal{A}$%
\end{tabular}
\right) \left(
\begin{tabular}{l}
$ X $ \\
$ Y $%
\end{tabular}
\right) =E \left(
\begin{tabular}{l}
$ X $ \\
$ Y $%
\end{tabular}
\right).
\end{equation}
Solutions of this set of linear equations yield the eigen energies
$E$ and the amplitudes $X,Y$ of the excited states. The dimension
of the matrices ${\mathcal{A}}, {\mathcal{B}}$ is the space size
of the two-quasiparticle configurations. Using  the finite rank
approximation the QRPA equations~(\ref{QRPA}) can be reduced to
the secular equation and the matrix dimensions never exceed $6N
\times 6N$ independently of the configuration space
size~\cite{svg08}.

We apply our approach to study characteristics of the low-lying
$2^+$ states in the even-even nuclei around $^{132}$Sn. We use the
Skyrme interaction SLy4~\cite{sly4} in the particle-hole channel
together with the isospin-invariant pairing force~(\ref{pair}).
The pairing strength $V_{0}$ is taken equal to -940 MeVfm$^{3}$ in
connection with the soft cutoff at 10 MeV above the Fermi energies
as introduced in Ref.~\cite{svg08}.
%
%
\section{Properties of low-lying quadrupole states}
We study the $2^+_1$ state energies and the $B(E2\uparrow)$-values
in $^{126-130}$Pd, $^{124-132}$Cd, $^{126-134}$Sn, $^{128-136}$Te,
$^{134-138}$Xe. The results of our calculations~\cite{svg08,svg09}
and the available experimental
data~\cite{Rad02,Jak02,Rad05,Jun07,cd128,Ram01} are shown in
Fig.~1 and Fig.~2. One can see that there is a correct
description of the isotopic and isotonic dependences. The $2^+_1$
energies have a maximal value at $N=82$ for the isotopes and at
$Z=50$ for the isotones. Such a behavior corresponds to a standard
evolution of the energies near closed shells. The structural
peculiarities are reflected in the  evolution of the $B(E2)$ values. The
$B(E2)$-value at $N=82$ for the isotopes ($Z=50$ for the isotones)
is either a maximal value in the Sn isotopes (the $N=82$
isotones), or a minimal value in the Pd, Cd, Te, Xe isotopes (the
$N=80,84$ isotones). As it was explained in
Ref.~\cite{svg08,svg09} the behaviour of the
$B(E2\uparrow)$-values is related with the proportion between the
amplitudes $X,Y$ for neutrons and protons. There is some
overestimation in our calculations in comparison with the
available experimental data. One can expect an improvement if the
coupling with the two-phonon components of the wave
functions~\cite{svg04} is taken into account.

%
\begin{table}[]
\caption[]{Structure of the $2_{1,4}^{+}$ states in $^{132}$Te.}
\begin{center}
\begin{tabular}{ccccc}
\hline\noalign{\smallskip}
          state &$\{n_{1}l_{1}j_{1}, n_{2}l_{2}j_{2}\}_{\tau}$& $X$ & $Y$ & structure[$\%$] \\
\hline\noalign{\smallskip}
 $2_{1}^{+}$    & $\{2d_{5/2},2d_{5/2}\}_{p}$  & 0.72 & 0.12 & 25\\
                & $\{1g_{7/2},2d_{5/2}\}_{p}$  & 0.30 & 0.05 &  8\\
                & $\{1g_{7/2},1g_{7/2}\}_{p}$  & 0.89 & 0.15 & 39\\
                & $\{1h_{11/2},1h_{11/2}\}_{n}$& 0.55 & 0.20 & 13\\
\\
 $2_{4}^{+}$    & $\{2d_{5/2},2d_{5/2}\}_{p}$  &-0.40 & 0.06 &  8\\
                & $\{1g_{7/2},2d_{5/2}\}_{p}$  &-0.16 & 0.02 &  3\\
                & $\{1g_{7/2},1g_{7/2}\}_{p}$  &-0.54 & 0.08 & 14\\
                & $\{1h_{11/2},1h_{11/2}\}_{n}$& 1.12 & 0.04 & 63\\
\noalign{\smallskip}\hline
\end{tabular}
\end{center}
\end{table}
%
Let us now discuss the structure of the $2^+_1$ state of
$^{124-132}$Cd and $^{126-130}$Pd. As shown in~\cite{svg09}, in
$^{124-132}$Cd the proton phonon amplitudes are dominant and the
contribution of the main proton configuration
$\{1g_{9/2},1g_{9/2}\}$ increases from 79\% in $^{124}$Cd to 89\%
in $^{128}$Cd, while the main neutron configuration
$\{1h_{11/2},1h_{11/2}\}$ exhausts about 13\%, 11\% and 7\% of the
wave function normalization in $^{124}$Cd, $^{126}$Cd and
$^{128}$Cd, respectively. The closure of the neutron subshell
$1h_{11/2}$ in $^{130}$Cd leads to the vanishing of neutron
pairing and as a result the energy of the first neutron
two-quasiparticle configuration $\{2f_{7/2},1h_{11/2}\}$ in
$^{130}$Cd is larger than energies of the first neutron
configurations in $^{128,132}$Cd. It follows that in $^{130}$Cd
the leading contribution (about 97\%) comes from the proton
configuration $\{1g_{9/2},1g_{9/2}\}$ and the $B(E2)$-value is
reduced. The structure of the $2^+_1$ state of $^{126,128,130}$Pd
is similar to that of $^{128,130,132}$Cd. We obtain a
noncollective structure with the dominance of the proton
configuration $\{1g_{9/2},1g_{9/2}\}$. In $^{128}$Pd, as it is
discussed for $^{130}$Cd, the contribution of the proton
$\{1g_{9/2},1g_{9/2}\}$ increases to 96\%. This again results in a
reduction of the $B(E2)$ value.

The dominance of the neutron-proton attraction is one of the main
feature of the effective nucleon-nucleon interaction. On the other
hand, the collective quadrupole isovector valence-shell
excitations, so-called mixed-symmetry states~\cite{pbl08,iach84},
are very sensitive to the proton-neutron interaction. That is why
it is interesting to study the characteristics of the lowest
isovector-collective states in nuclei around $^{132}$Sn. As an
example we consider $^{132}$Te. The results of our calculation are
discussed in more detail in Ref.~\cite{mixsym}. Table 1 shows that
the dominant neutron and proton phonon amplitudes $X,Y$ of the
collective $2^+_1$ states are in phase and this corresponds to the
typical isoscalar character of the lowest-lying quadrupole
excitations. The next fairly collective state is the $2^+_4$
state. The calculated energy of the $2^+_4$ state is equal to
$3.1$~MeV with a calculated excitation strength of
$B(E2;0^+_{gs}\rightarrow2^+_4)=330 \ e^{2}$fm$^{4}$. The dominant
neutron and proton amplitudes of the $2^+_4$ state are in opposite
phase. The isovector character of the $2^+_4$ is reflected in the
noticeable size of the $B(M1;2^+_4\rightarrow 2^+_1)$ value equal
to 0.25$\mu_{N}^{2}$. This analysis can help to identify the
mixed-symmetry states in $^{132}$Te, but it is only a rough
estimate since the effects of the phonon-phonon coupling are not
considered.
%
%
\section{Conclusions}
A finite rank separable approximation for the QRPA calculations
with Skyrme-type interactions is presented. This approach enables
one to reduce remarkably the dimensions of the matrices that must
be diagonalized to perform structure calculations in very large
configuration spaces. Using the same set of parameters we have
investigated the evolution of the $2^+_1$ state energies and the
$B(E2)$-values in $^{126-130}$Pd, $^{124-132}$Cd, $^{126-134}$Sn,
$^{128-136}$Te, $^{134-138}$Xe. Our calculations for the energies
and the $B(E2)$-values describe correctly the isotopic and
isotonic dependences. We give predictions for the structure of the
$2^+_1$ state in $^{126-130}$Pd, $^{124-132}$Cd. As an
illustration of the method we study the lowest isovector
collective quadrupole state in $^{132}$Te. It is found from our
calculations that the $2^+_4$ state might be the best candidate
for the mixed-symmetry state. To finalize the study one needs to
take into account the phonon-phonon coupling.
%
%
\section*{Acknowledgments}
We are grateful to Prof.~R.~V.~Jolos and Prof.~Ch.~Stoyanov for
useful discussions. This work was supported in parts by the
Heisenberg-Landau program and by the IN2P3-JINR agreement.
%
%

%
\begin{figure*}[t!]
\includegraphics[width=15cm]{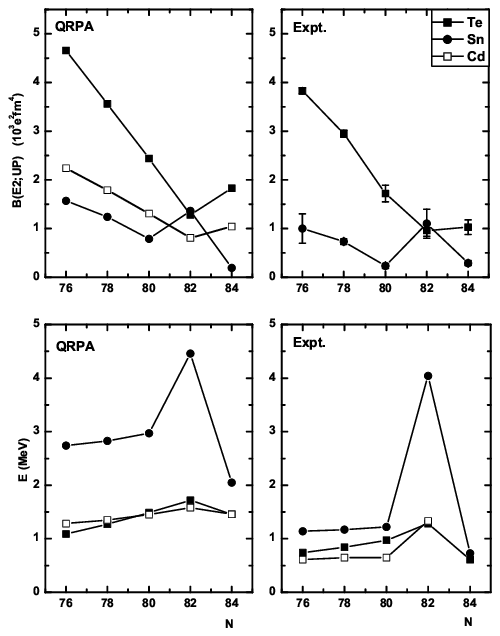}
\caption { Energies and $B(E2; 0^+_{gs}\rightarrow 2^+_1)$ values
of $2^+_1$ states in $^{124-132}$Cd, $^{126-134}$Sn,
$^{128-136}$Te. Results of the QRPA calculations are represented
in the left panels; the available experimental data are in the
right panels.}
\end{figure*}
%
%
\begin{figure*}[t!]
\includegraphics[width=15cm]{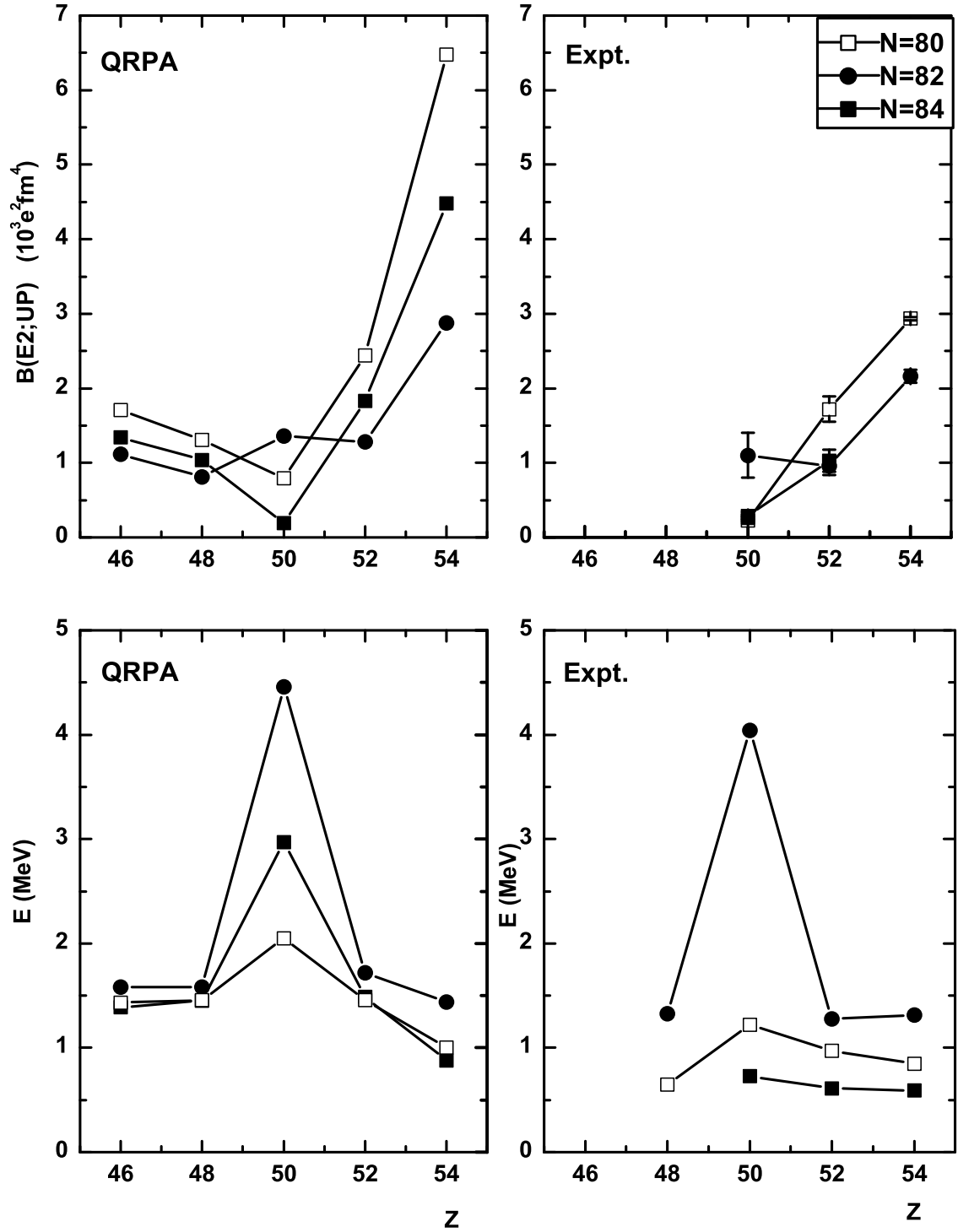}
\caption {Energies and $B(E2; 0^+_{gs}\rightarrow 2^+_1)$ values
of $2^+_1$ states in the $N=80,82,84$ isotones. Results of the
QRPA calculations are represented in the left panels; the
available experimental data are in the right panels.}
\end{figure*}
%
\end{document}